# Understanding Optical Anisotropy in Multilayer γ-InSe and ε-GaSe

Jason Lynch[1,†], Zexuan Liu[1,†], Sergiy Krylyuk[2], Huairuo Zhang[2,3], Albert Davydov[2], Deep Jariwala[1,*]

[1]Electrical and Systems Engineering, University of Pennsylvania, Philadelphia, PA 19104, USA

[2]Materials Science and Engineering Division, National Institute of Standards and Technology, Gaithersburg, MD 20899, USA

[3]Theiss Research, Inc., La Jolla, CA 92037, USA

[†]Authors contributed equally

[*]Corresponding Author: dmj@seas.upenn.edu

**Abstract**

Low-dimensional media have exhibited optical anisotropy that is unachievable in traditional 3D media due to the asymmetry of their strong, in-plane covalent bonds and weak out-of-plane van der Waals interactions. As a result, 2D media are promising building blocks for ultrathin devices such as polarimeters, polarized light sources, and active polarizers. III-VI semiconductors possess a rare property in the class of multilayered semiconductors, which is that their fundamental excitons are oriented out-of-plane. This allows them to exhibit phenomena such as transparency in the visible range while also being emissive in the visible and near-infrared ranges. Here, we report the first experimental values for the anisotropic refractive indices of γ-InSe and ε-GaSe, and we observe the effects of the out-of-plane excitons on the c-axis refractive index. It is found that both materials exhibit moderate optical anisotropy for multilayered semiconductors. The complex, anisotropic refractive index of γ-InSe and ε-GaSe enables the accurate simulation of these media, allowing for the design of high-performance, ultra-compact optoelectronic devices.

**Introduction**

Optical anisotropy leads to polarization-dependent light-matter interactions enabling phenomena such as the spatial separation of polarized light, waveplates, and improved polarized photodetectors[1–3]. Optical anisotropy occurs in systems where symmetry is broken along one or more directions, and it results in the medium exhibiting different susceptibilities depending on the direction of the electric field[4]. In systems where there is only one asymmetric axis, such as the systems we study here, the medium is said to be uniaxial. In uniaxial systems, the asymmetric axis is referred to as the extraordinary axis and the other axes are the ordinary axes. When the electric field of incident light is parallel and perpendicular to the extraordinary axis, its interaction with the medium is determined by the extraordinary and ordinary refractive indices, respectively.

The study of optical anisotropy has largely focused on bulk media such as liquid crystals[5], calcite[6], and LiNbO$_3$[7] since these media are lossless in the visible range and they can be integrated into photonic devices at the commercial scale. However, the birefringence values, i.e., the difference between the extraordinary and ordinary refractive indices, in these media are less than 0.2. Due to their relatively modest birefringence, bulk media typically require long interaction lengths, which prevents their use in ultra-compact optical devices. Recently, low-dimensional, multilayered materials have been shown to exhibit above-unity birefringence due to the asymmetry of the in-plane covalent bonds and out-of-plane van der Waals interactions[8–11]. In addition to their ability to produce nanometer-thin films, low-dimensional semiconductors also have the advantage of improved flexibility, and they host excitons that are both highly-emissive and highly-tunable[12–15]. Further, advances in the growth of 2D materials have allowed for their growth on the wafer scale[16].

III-VI multilayered semiconductors, specifically InSe and GaSe, are a widely studied subclass of 2D semiconductors that possess excellent electrical properties, including high carrier mobility[17], ideal optical properties, including emissivity in the bulk[18], and lack of centrosymmetry, which leads to both piezo- and ferroelectricity[19]. A unique property of InSe and GaSe compared to other multilayered semiconductors is that their primary excitons are oriented out-of-plane due to the symmetries of the conduction band minima and valence band maxima[20–22]. This orientation allows excitons in InSe and GaSe to more easily couple to surface-plasmon-polaritons and be more sensitive to out-of-plane stimuli[23,24]. However, the out-of-plane refractive indices associated with these excitons are yet to be experimentally determined, which prevents the modelling of optoelectronic devices that can take advantage of their unique properties.

Here, we report the anisotropic refractive indices in III-VI multilayered semiconductors, γ-InSe and ε-GaSe, using spectroscopic ellipsometry from the deep ultraviolet to the near infrared (λ = 190 nm to 1000 nm). We observe the out-of-plane excitons in both γ-InSe and ε-GaSe in the out-of-plane refractive indices. We also observe that both media exhibit larger birefringence than classical, bulk 3D media. However, the birefringence and dichroism of γ-InSe and ε-GaSe are smaller than other 2D media such as MoS$_2$. We mainly attribute the lower optical anisotropy to the vertical orientation of the metal-metal bonds, reducing the structural anisotropy and a smaller difference in optical band gap along the ab- and c-axes in the III-VI semiconductors compared to MoS$_2$.

# Results & Discussion

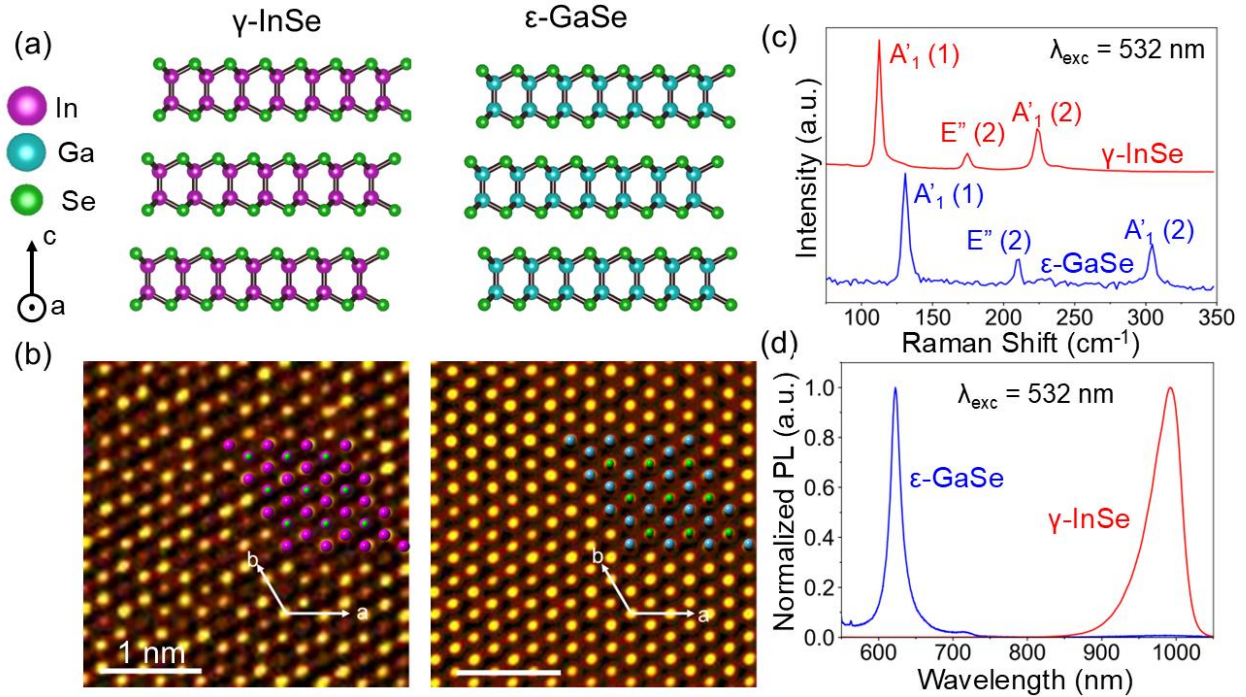

**Figure 1. Structural characterization of γ-InSe and ε-GaSe. (a)** Side view diagram of the crystal structure of γ-InSe (left) and ε-GaSe (right) showing the different stacking pattern of the γ- and ε-polytypes. **(b)** Plan-view ADF-STEM images of γ-InSe (left) and ε-GaSe (right) with atomic model overlays confirming each crystals polytype. Scale bars are 1 nm. Normal incidence **(c)** Raman and **(d)** PL spectroscopy of γ-InSe (red) and ε-GaSe (blue) showing their excellent optical properties using an excitation wavelength ($\lambda_{exc}$) of 532 nm. The PL peak of γ-InSe appears asymmetric due to the sensitivity of the detector rapidly decreasing at longer wavelengths. Although the excitons in both crystals are mainly aligned along the c-axis, their emission is observed with normal incidence measurements because the 0.35 numerical aperture of the lens causes the crystals to be excited by light with incident angles from 0° to 20.5°.

Bulk crystals of γ-InSe and ε-GaSe are grown by the vertical Bridgman method (See Methods). This process has been shown to produce high-quality crystals of both γ-InSe and ε-GaSe[25–28]. Individual layers of both ε-InSe and γ-GaSe are hexagonal, and they consist of two monoatomic sheets of the metal (In or Ga) sandwiched between two sheets of Se[29]. Each layer is dangling-bond-free, and each metal atom is covalently bonded to three Se atoms and another metal atom. However, the ε- and γ-polytypes differ by how these layers are stacked along the out-of-plane c-axis. In the γ-polytype, each layer is shifted by 1/3 of the unit cell along the <10$\bar{1}$0> direction, and the crystal, and the crystal is in the R3m space group (Figure 1a, left)[29]. As for the ε-polytype, alternating layers are shifted by 2/3 and 1/3 of the until cell along the <1010> direction (a-axis), and the crystal belongs to the $P\bar{6}m2$ space group, which lacks an inversion center (Figure 1a, right)[29]. The crystal structure of γ-InSe has been confirmed by cross-sctional annular dark field scanning stransmission electron microscopy (ADF-STEM) in our previous works [25,30].

The ε-GaSe structure is confirmed with X-ray diffraction measurements (Figure S1). Figure 1b shows the atomic resolution plan-view HAADF-STEM images of the γ-InSe (left) and ε-GaSe (right), respectively, which are perfectly matched with the overlaid atomic models of the polytypes.

The crystals are exfoliated onto $SiO_2$/Si and c-plane sapphire substrates for characterization (See Methods). Their high quality is further confirmed using Raman spectroscopy (Figure 1c). The A'$_1$ (1), E" (2), and A'$_1$ (2) peaks in γ-InSe occur at 114 cm$^{-1}$, 176 cm$^{-1}$, and 225 cm$^{-1}$, respectively, which only differ from the literature values by 2 cm$^{-1}$, 1 cm$^{-1}$, and 1 cm$^{-1}$, respectively[31,32]. For ε-GaSe, the A'$_1$ (1), E" (2), and A'$_1$ (2) peaks occur at 132 cm$^{-1}$, 210 cm$^{-1}$, and 306 cm$^{-1}$, respectively, which all differ from literature by 2 cm$^{-1}$ [32,33]. Further, both γ-InSe and ε-GaSe have direct bandgap in the bulk at the Γ-point with energies of 1.29 eV and 2 eV, respectively[34,35]. We confirm these properties and the high optical quality of our crystals using normal incidence photoluminescence (PL) spectroscopy (Figure 1d). As mentioned above, both γ-InSe and ε-GaSe have out-of-plane excitons due to their optical selection rules, which is an uncommon property in multilayered semiconductors[20–22]. However, we still observe PL from the crystals because the numerical aperture of the lens (0.35) causes the incident angle of light to be between 0° and 20.5°. Therefore, the normal incidence PL measurement can still excite the excitons since there will be a vertical electric field component of the incident light. The PL peak of γ-InSe appears to be asymmetric because the efficiency of the Si detector rapidly decreases with wavelength in the spectral range of the PL peak (See Methods).

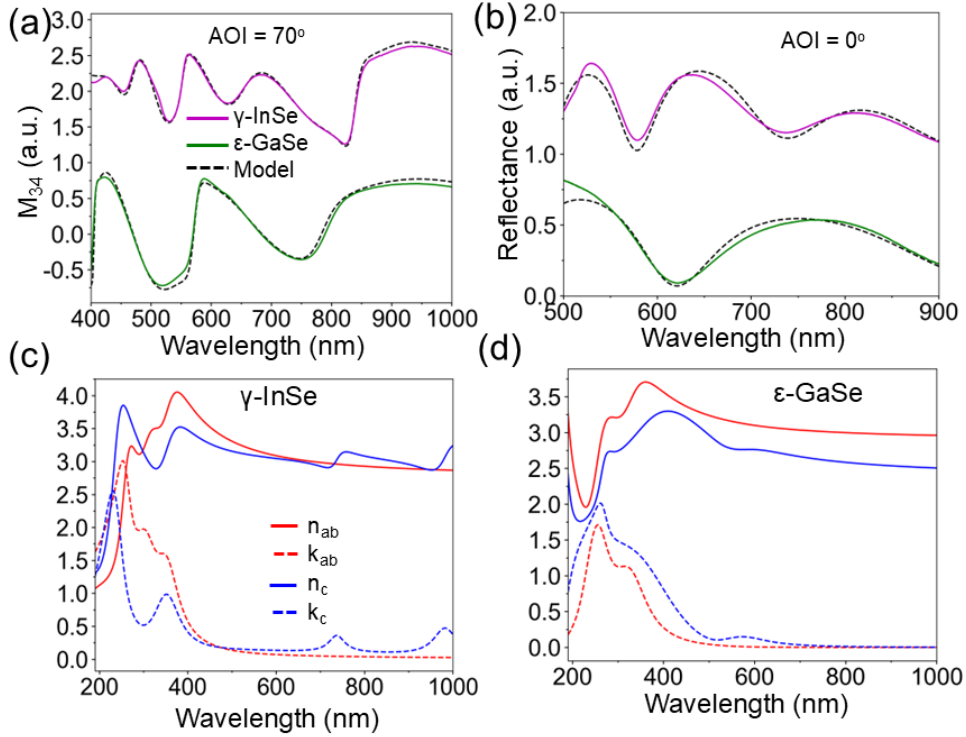

**Figure 2. Anisotropic optical constants. (a)** Anisotropic element of the Mueller matrix ($M_{34}$) of γ-InSe (magenta) and ε-GaSe (green) at an angle of incidence (AOI) of 70°. Due to detector limitations, standard spectroscopic ellipsometry is performed from λ = 190 nm to 400 nm (Figure S4) while Mueller matrix ellipsometry (solid) is performed from λ = 400 nm to 1000 nm (Figure S5). The model (black dashed lines) reproduced the experimental results with an RMSE value of 4.68 and 4.31 for γ-InSe and ε-GaSe, respectively. **(b)** The experimental normal incidence reflectance and model calculated using the transfer matrix method. The γ-InSe data in (a) and (b) have been shifted upwards for clarity. The real (solid lines) and imaginary (dashed lines) parts of the refractive indices of **(c)** γ-InSe and **(d)** ε-GaSe. The extraordinary axes (blue lines) in both crystals are along the c-axis, while the ordinary axes (red lines) are along the a- and b-axes.

Spectroscopic ellipsometry is a powerful optical characterization technique as it uses information from both the amplitude and phase of reflected light to determine the complex refractive index of a medium (Section S1)[36]. The layered structure of γ-InSe and ε-GaSe results in both crystals displaying optical anisotropy with the extraordinary axis along the c-axis. The anisotropic refractive index of the crystals can be measured by either performing standard spectroscopic ellipsometry (SSE) over a large range of incident angles of light[8], or by performing Mueller matrix ellipsometry (MME) with improved accuracy when done over a large range of incident angles[10]. Here, we first perform SSE on the SiO$_2$/Si substrate to improve the accuracy of the model (Figure S2). The anisotropic complex of refractive indices of γ-InSe and ε-GaSe are then determined by performing ellipsometry from 190 nm to 1000 nm at angles of incidence (AOIs) of 50°, 60°, and 70° (Figures 2a and S3-S4). Due to detector limitations, MME is performed over the wavelength range of 400 nm to 1000 nm, while SSE is performed from 190 nm to 400 nm (See Methods). Furthermore, to improve the accuracy of the models, our fitting process used data on c-plane sapphire for the SSE and data on SiO$_2$/Si for the MME (See Methods). This was done since the high refractive index and low loss of Si in the visible increases the signal-to-noise ratio of the measurement in the visible and near IR. However, Si is absorptive in the UV, making the sapphire substrate more well-suited for the measurement. The optical constants of the crystals are both modelled using a multi-Tauc-Lorentz model where each Tauc-Lorentz oscillator represents a band-to-band optical transition (Section S1)[36]. The DataStudio software from Park Systems is used to develop a model that closely follows the experimental results. The root-mean-squared-error (RMSE) characterizes the accuracy of the model, and the RMSE is found to be 4.68 and 4.31 for γ-InSe and ε-GaSe, respectively, which is below the typically desired RMSE of 5.[36] We further confirm the accuracy of the model by measuring the normal incidence reflectance on SiO$_2$/Si (Figure 2b). The reflectance dips are cavity modes caused by the wavelength-scale thickness of the flakes (284 nm and 103 nm for γ-InSe and ε-GaSe, respectively). The presence of these modes confirms that the flakes are transparent to normal-incidence light in the visible range despite being emissive, which is a rare property for a multilayered semiconductor[18]. This phenomenon is caused by the out-of-plane orientation of the excitons in γ-InSe and ε-GaSe allows for the in-plane complex refractive index to be lossless.

This agrees with our measurement of the anisotropic complex refractive index of γ-InSe and ε-GaSe (Figure 2c-d). Both media are nearly lossless in the ab-axes while also displaying out-of-plane oriented exciton resonances as predicted by theory[20]. The lowest energy in-plane resonances in γ-InSe and ε-GaSe occur at 352 nm and 320 nm, respectively, which are both in the near-UV range. At longer wavelengths, the c-axis of γ-InSe shows an exciton at 737 nm, and its fundamental exciton at 982 nm. The location of the A-exciton agrees with the observed PL peak at 992 nm with the slight redshift being attributed to the exciton binding energy[37]. In accordance with the Kramers-Konig relation and Tauc-Lorentz model, the maximum refractive index occurs at a longer wavelength than the exciton resonance. As a result, the maximum refractive index due to the primary exciton occurs outside of our measurement range. Still, we are able to clearly observe the c-axis exciton in $k_c$ graph. ε-GaSe also shows a c-axis-oriented exciton at 604 nm compared to its PL peak at 622 nm.

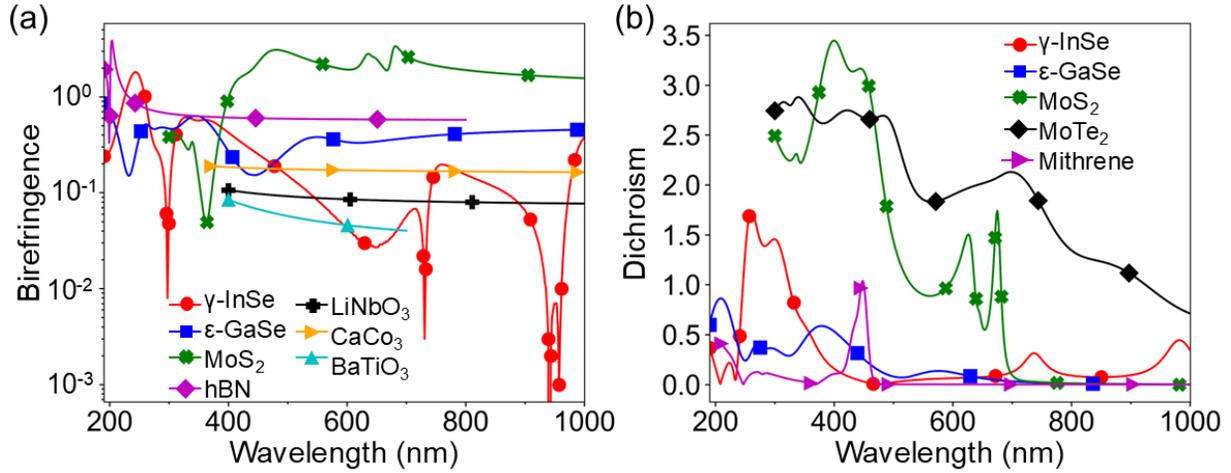

**Figure 3. Comparison of optical anisotropy. (a)** Birefringence ($|n_{ab}-n_c|$) and **(b)** dichroism ($|k_{ab}-k_c|$) in γ-InSe and ε-GaSe with other semiconductors and insulators from literature[7,8,38–40]. LiNbO$_3$, CaCO$_3$, and BaTiO$_3$ are all three-dimensional crystals, mithrene is a layered inorganic-organic medium, and γ-InSe, ε-GaSe, MoS$_2$, MoTe$_2$, and hBN are all multilayered inorganic media. The sharp drops in birefringence in γ-InSe indicate polarization singularities ($n_{ab} = n_c$) which occur near the emissive and sensitive excitons.

ε-GaSe exhibits a birefringence ($|n_{ab}-n_c|$) that averages 0.36 in the visible and near infrared (λ = 400 nm to 1000 nm) while γ-InSe exhibits an average of 0.12 (Figure 3a). Both of these values are larger than those in classic 3D birefringent media (LiNbO$_3$, CaCO$_3$, and BaTiO$_3$)[6,7,38]. In addition to increased birefringence, γ-InSe and ε-GaSe have several advantages over these 3D media including the ability to fabricate atomically thick layers, light emission in the visible and near IR, tunable excitons, and high optical nonlinearities. However, it is important to mention that despite our flakes being transparent in the visible range, their absorption is still larger than the 3D media due to their large optical band gaps. γ-InSe and ε-GaSe have lower birefringence values than other multilayered media (MoS$_2$ and hBN)[10,40]. The giant optical anisotropy in multilayered 2D media is commonly attributed to the anisotropy in the in-plane and out-of-plane bonding and transitions[8]. For example, the in-plane direction of MoS$_2$ has covalent bonds and a strong, bright exciton. In contrast, its out-of-plane direction is dominated by van der Waals interactions between layers, and the lowest resonance in this direction is > 4 eV[10]. However, in γ-InSe and ε-GaSe, the bond between the metal atoms is oriented vertically, reducing its optical anisotropy, and the energy difference between the fundamental resonances along the ab- and c-axes is smaller than in MoS$_2$. This results in lower birefringence values in III-VI media than in other 2D media. This is further confirmed by comparing the dichroism ($|k_{ab}-k_c|$) of γ-InSe and ε-GaSe to other 2D media (Figure 3b). However, γ-InSe and ε-GaSe both possess unique properties compared to other 2D media. Most notably in terms of optical anisotropy, the vertical orientation of their excitons makes them highly sensitive to pressure changes through changes in their interlayer spacing and allows them to strongly couple to the vertical electric field of plasmon polaritons[18,23,24,41].

**Conclusion**

In conclusion, we have determined the anisotropic refractive index of γ-InSe and ε-GaSe using spectroscopic ellipsometry from λ = 190 nm to 1000 nm. The birefringence and dichroism of the III-VI semiconductors are compared to other 3D and 2D anisotropic media. It is found that γ-InSe and ε-GaSe is more anisotropic than traditional 3D media, but both flakes are less anisotropic than other 2D media such as Mo dichalcogenides. We attribute this to the reduced anisotropic both in the bonding, specifically due to the vertical metal-metal bonds, and a smaller difference in optical band gap along the ab- and c-axes compared to $MoS_2$. We also provided the first measurement of the refractive index associated with the out-of-plane excitons in these III-VI semiconductors. The out-of-plane exciton allows γ-InSe and ε-GaSe to be transparent to normal incident visible range despite being highly emissive in the visible and near infrared[20]. Therefore, the anisotropic refractive index provided here will enable the design of new, high-performance optoelectronic devices using γ-InSe and ε-GaSe.

**Methods**

*Crystal Growth and Exfoliation*

InSe and GaSe single crystals were grown by the vertical Bridgman method using polycrystalline $In_{1.04}Se_{0.96}$ and GaSe charges, respectively. The charges were vacuum-sealed in graphitized quartz ampoules. The InSe (GaSe) melt was equilibrated at 720 °C (990 °C) for several hours, and then the ampoule was translated across a temperature gradient at a rate of 0.5 mm/h (0.4 mm/h). The γ-phase of InSe and the ε-phase of GaSe were confirmed by X-ray diffraction and transmission electron microscopy measurements.

γ-InSe and ε-GaSe were mechanically exfoliated from bulk crystals using Scotch tape and transferred onto the $SiO_2$/Si and single-side polished sapphire (c-plane) substrates. The substrates were cleaned using acetone before and after the transfer to remove any contaminants.

*X-ray Diffraction*

Powder X-Ray Diffraction (PXRD) measurements were conducted using a Bruker D8 Advance diffractometer (Bruker AXS, Inc., Billerica, MA) equipped with an EIGER2 R 500 K (Dectris, Inc., Philadelphia, PA) single-photon-counting detector. PXRD data were acquired on finely ground InSe and GaSe powders using Cu K$\alpha$ ($\lambda$ = 0.15418 nm) radiation in the 2θ-angle range of 15° to 100°. The patterns were analyzed using commercial MDI Jade version 6.5 software to refine the unit cell parameters.

*Transmission Electron Microscopy*

Samples for plain-view scanning transmission electron microscopy (STEM) were exfoliated from the InSe and GaSe flake to take the [001] zone-axis data. An FEI Titan 80-300 probe-corrected STEM/TEM microscope operating at 300 keV was employed to conduct atomic resolution annular dark-field scanning transmission electron microscopy (ADF-STEM) imaging analysis, with a probe convergence semi-angle of 14 mrad and a collection angle of 34 mrad to 195 mrad.

*Raman, Reflectance, and Photoluminescent Spectroscopy*

Raman, normal incidence reflectance, and photoluminescence measurements were performed using a Horiba Scientific confocal microscope (LabRAM HR Evolution). This instrument used an Olympus objective len (×100) and two different grating -based spectrometers (100 lines/mm for reflectance/PL and 600 lines/mm for Raman), which are coupled to a Si focal plane array detector. A continuous-wave excitation source with excitation wavelength of 532 nm was utilized.

*Ellipsometry*

Imaging ellipsometry measurements for determining the dielectric function were performed using the Accurion EP4 system (Park Systems) over the 190 nm to 1000 nm spectral range with a 2 nm step size. A 7× objective lens was used for all measurements. Multi-angle incidence measurements were conducted over an angular range of 50° to 70°. The measured data were analyzed using the EP4 model and DataStudio software provided by the EP4 system. The DataStudio software allows us to simultaneously fit the SSE data for the flakes on sapphire and the MME data for the flakes on $SiO_2$/Si with the fit parameters being the refractive index of the flakes, the thickness of the flakes on each substrate, and the surface roughness of the flakes on each substrate. The surface roughness is modelled using the Bruggman effective medium approximation between the in-plane refractive index of the crystals and air.

**Acknowledgements**

D.J. and J.L. acknowledge support from the Office of Naval Research Young Investigator Award (N00014-23-1-2037), Metamaterials program and the National Science Foundation, Future of Semiconductors (FuSe) program award number 2328743. This work was carried out in part at the Singh Center for Nanotechnology at the University of Pennsylvania, which is supported by the National Science Foundation National Nanotechnology Coordinated Infrastructure Program under grant NNCI-2025608. H.Z. acknowledges support from NIST under the financial assistance award 70NANB22H101.

Disclaimer: Certain commercial equipment, instruments, software, or materials are identified in this document to specify the experimental procedure adequately. Such identification is not intended to imply recommendation or endorsement by the National

Institute of Standards and Technology, nor is it intended to imply that the materials or equipment identified are necessarily the best available for the purpose.

# Supporting Information: Understanding Optical Anisotropy in Multilayer γ-InSe and ε-GaSe

Jason Lynch[1,†], Zexuan Liu[1, †], Sergiy Krylyuk[2], Huairuo Zhang[2,3], Albert Davydov[2], Deep Jariwala[1,*]

[1]Electrical and Systems Engineering, University of Pennsylvania, Philadelphia, PA 19104, USA

[2]Materials Science and Engineering Division, National Institute of Standards and Technology, Gaithersburg, MD 20899, USA

[3]Theiss Research, Inc., La Jolla, CA 92037, USA

[†]Authors contributed equally

[*]Corresponding Author: dmj@seas.upenn.edu

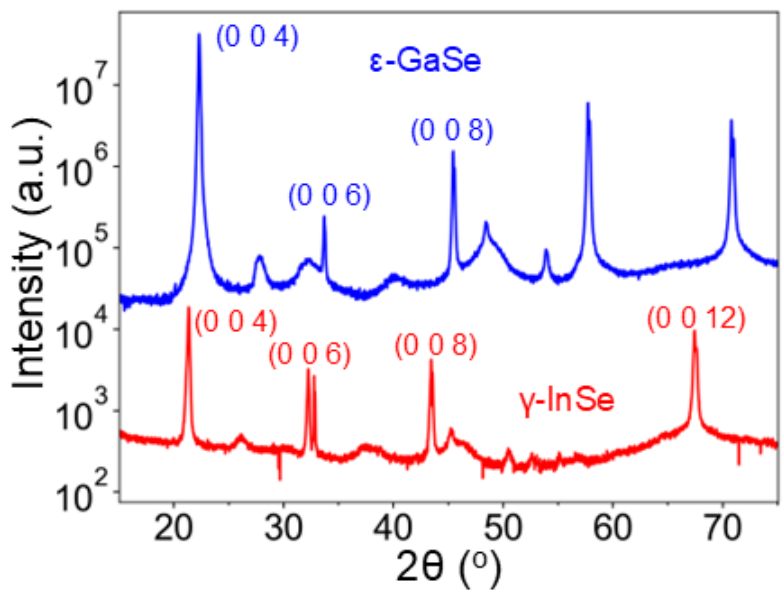

**Figure S1. X-Ray diffraction measurement.** The X-ray diffraction of γ-InSe (red) and ε-GaSe (blue) with some peaks labelled showing the high-quality of the bulk crystals that are exfoliated in this work.

**Section S1: Spectroscopic Ellipsometry and Modelling**

Standard spectroscopic ellipsometry (SSE) uses the difference in complex reflection coefficients ($\tilde{r} = |r|e^{i\phi}$) between transverse magnetic (TM) and transverse electric (TE) polarized light at non-normal angles of incidence to determine the complex permittivity of films[1]. SSE does this by characterizing the ratio of reflected amplitude (ψ = arctan(|$r_{TM}$|/|$r_{TE}$|)) and the phase difference between the two polarizations (Δ = $\Phi_{TM} - \Phi_{TE}$). SSE can be used to measure the anisotropic refractive index in γ-InSe and ε-GaSe by performing the measurement over a range of angles of incidence (AOI). However, Mueller matrix ellipsometry (MME) is a more well-suited characterization technique in this case.

The Mueller matrix (MM) describes the transformation of the Stokes vector of incident light upon reflection[1]. The Stokes vector has 4 elements where the first element is commonly normalized to 1, and the second, third, and fourth elements are the difference in reflected intensity of TM and TE, 45° and -45°, and circular left-hand and right-hand polarized light, respectively. The second through fourth elements are also the principle axes of the Poincare Sphere so the Stokes vector presents a complete description of the polarization of light. Therefore, the MM provides a complete description of the transformation of the polarization of light, and it can be used to accurately measure the anisotropic refractive index of a medium using a single AOI. Although this can be done using a single angle of incidence, we performed MME at 50°, 60°, and 70° for improved accuracy.

The refractive index is then retrieved by constructing a model that recreates the experimental data. As γ-InSe and ε-GaSe are both semiconductors, their optical constants can be modelled using Tauc-Lorentz oscillators[1]:

$$\varepsilon_2(E) = \begin{cases} \frac{(E-E_g)^2}{E^2} \frac{fE_x\Gamma E}{(E_x^2 - E^2)^2 + \Gamma^2 E^2} &, E > E_g \\ 0, E \leq E_g \end{cases}$$

where $\varepsilon_2$ is the imaginary part of the permittivity, E is the energy of incident light, $E_g$ is the Tauc gap (oscillator is lossless below this energy), f is the oscillator strength, and Γ is the linewidth. The real part of the permittivity can then be obtained from $\varepsilon_2$ using the Kramer-Kronig relation. The permittivity (ε = (n + ik)$^2$) along the ordinary and extraordinary axes of the crystals can then be modelled as a sum of Tauc-Lorentz oscillators ($\varepsilon_{TL}$) and a background permittivity ($\varepsilon_\infty$) where each oscillator represents an interband transition:

$$\varepsilon(E) = \varepsilon_\infty + \sum_i \varepsilon_{TL,i}(E)$$

The fit parameters of the model, along with the thickness of the layered crystals, are then varied to minimize the weighted root-mean-squared-error (RMSE) using the DataStudio software from Park systems. The RMSE is calculated as

$$RMSE = \sqrt{\frac{1}{2p_1 - q}\sum_i \left((\psi_i^{mod} - \psi_i^{exp})^2 + (\Delta_i^{mod} - \Delta_i^{exp})^2\right) + \frac{1}{11p_2 - q}\sum_{i,j}\left(M_{ij}^{mod} - M_{ij}^{exp}\right)^2}$$

where $p_1$ and $p_2$ refer to the number of wavelengths measured during the SSE and MME measurements, respectively, q is the number of fit parameters, and the superscripts "mod" and "exp" denote the modeled and experimental values, respectively. Typically, an RMSE less than 5 is considered accurate. The factor of two in front of $p_1$ is due to two experimental values being measured at each wavelength. As for $p_2$, the factor of 11 is because only the first three rows of the MM is measured, and $M_{11}$ is normalized to one. Therefore, 11 values of the MM are measured at each wavelength.

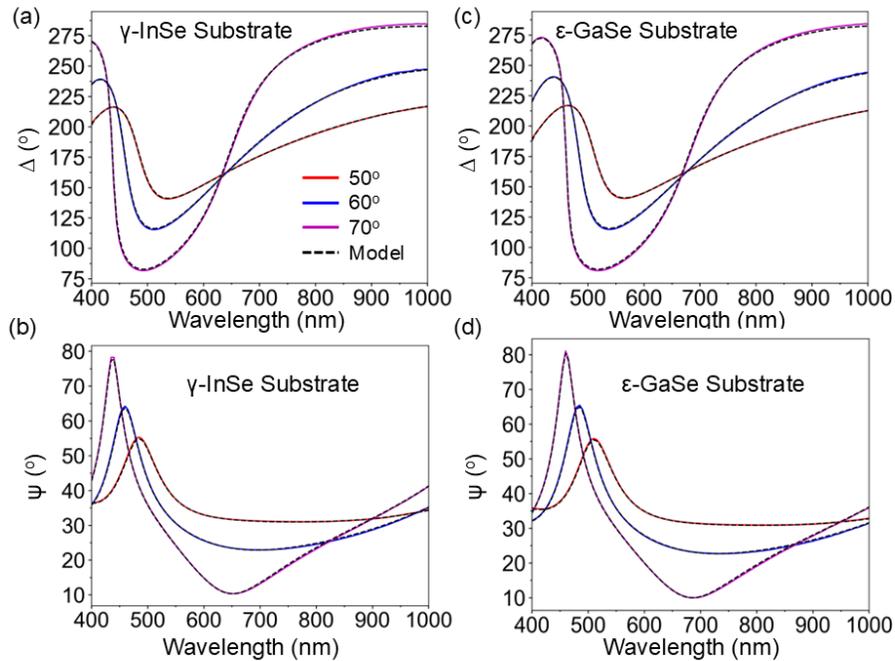

**Figure S2. Ellipsometry of SiO$_2$/Si substrates. (a, b)** Δ and ψ of the SiO$_2$/Si substrate that γ-InSe is exfoliated and **(c, d)** Δ and ψ of the SiO$_2$/Si substrate that ε-GaSe is exfoliated at incident angles of 50° (red), 60° (blue), and 70° (magenta). Performing ellipsometry on the substrate improves the accuracy of our models for γ-InSe and ε-GaSe since it determines the thickness of the SiO$_2$ layer. The SiO$_2$ is found to be 290 nm and 307 nm thick for the substrates of the γ-InSe and ε-GaSe, respectively.

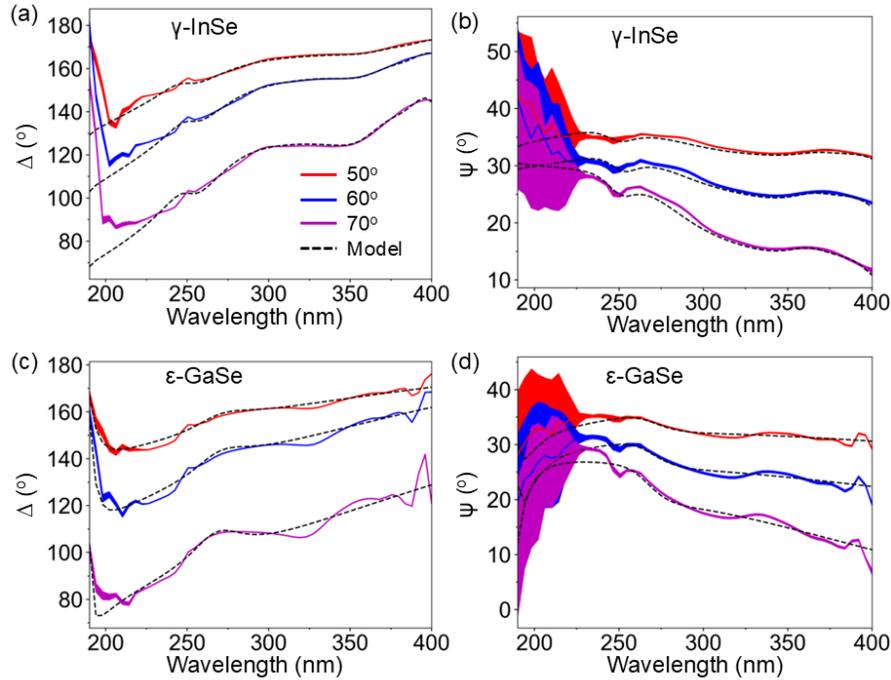

**Figure S3. SSE of γ-InSe and ε-GaSe on c-plane sapphire.** SSE data (Δ and ψ) of **(a-b)** γ-InSe and **(c-d)** ε-GaSe on a c-plane sapphire substrate at AOIs of 50° (red), 60° (blue), and 70° (magenta). The model (black dashed) developed by the DataStudio software shows good agreement with the experimental data. RMSE over the full spectral range (λ = 190 nm to 1,000 nm) is 4.68 and 4.31 for γ-InSe and ε-GaSe, respectively.

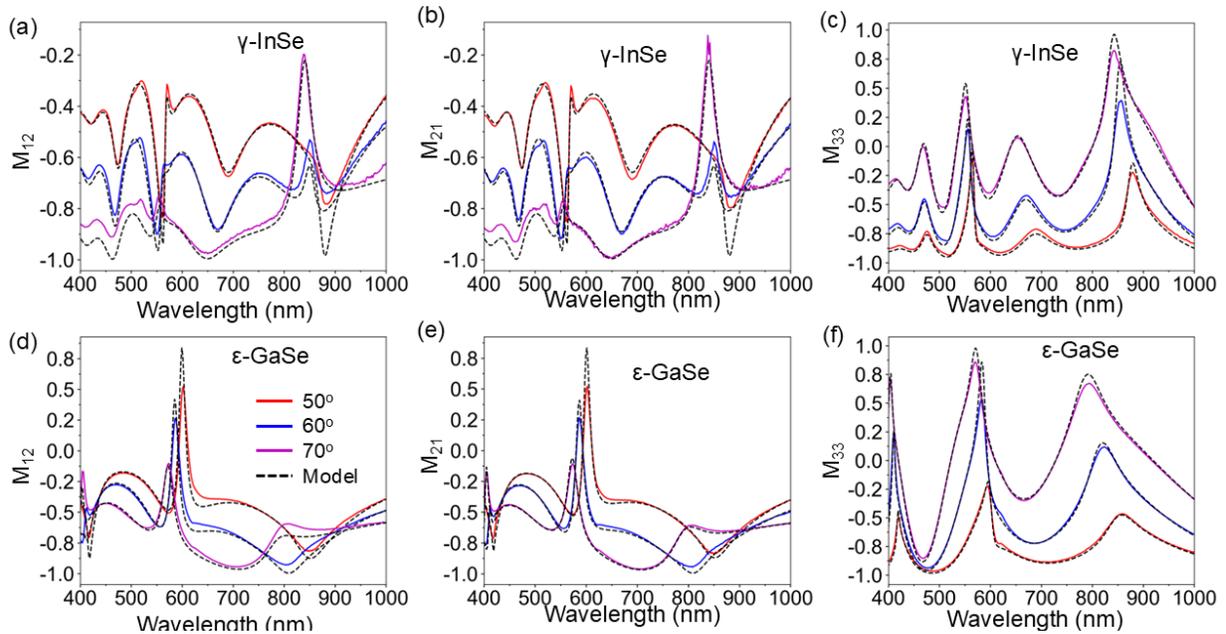

**Figure S4. MME of γ-InSe and ε-GaSe.** Select MM elements ($M_{12}$, $M_{21}$, and $M_{33}$) of **(a-c)** γ-InSe and **(d-f)** ε-GaSe on a SiO$_2$/Si substrate at AOIs of 50° (red), 60° (blue), and 70° (magenta). The model (black dashed)

developed by the DataStudio software shows good agreement with the experimental data. RMSE over the full spectral range (λ = 190 nm to 1,000 nm) is 4.68 and 4.31 for γ-InSe and ε-GaSe, respectively.

**Table S1. Complex, anisotropic refractive index of γ-InSe and ε-GaSe.**

| Wavelength (nm) | γ-InSe | | | | ε-GaSe | | | |
|---|---|---|---|---|---|---|---|---|
| | $n_{ab}$ | $k_{ab}$ | $n_c$ | $k_c$ | $n_{ab}$ | $k_{ab}$ | $n_c$ | $k_c$ |
| 190 | 1.074 | 1.623 | 1.315 | 1.255 | 3.275 | 0.175 | 2.423 | 0.773 |
| 192.025 | 1.084 | 1.65 | 1.323 | 1.32 | 3.092 | 0.197 | 2.262 | 0.849 |
| 194.05 | 1.094 | 1.678 | 1.335 | 1.387 | 2.942 | 0.219 | 2.139 | 0.92 |
| 196.075 | 1.104 | 1.706 | 1.351 | 1.456 | 2.815 | 0.243 | 2.045 | 0.986 |
| 198.1 | 1.115 | 1.736 | 1.372 | 1.528 | 2.706 | 0.269 | 1.971 | 1.047 |
| 200.125 | 1.125 | 1.766 | 1.397 | 1.601 | 2.609 | 0.296 | 1.914 | 1.104 |
| 202.15 | 1.135 | 1.798 | 1.427 | 1.676 | 2.523 | 0.326 | 1.87 | 1.157 |
| 204.175 | 1.146 | 1.831 | 1.462 | 1.752 | 2.445 | 0.358 | 1.836 | 1.206 |
| 206.2 | 1.157 | 1.866 | 1.504 | 1.83 | 2.374 | 0.392 | 1.81 | 1.252 |
| 208.225 | 1.169 | 1.902 | 1.552 | 1.908 | 2.309 | 0.43 | 1.791 | 1.295 |
| 210.25 | 1.181 | 1.94 | 1.608 | 1.988 | 2.249 | 0.471 | 1.778 | 1.334 |
| 212.275 | 1.195 | 1.979 | 1.672 | 2.067 | 2.195 | 0.516 | 1.768 | 1.372 |
| 214.3 | 1.209 | 2.021 | 1.745 | 2.145 | 2.145 | 0.564 | 1.763 | 1.407 |
| 216.325 | 1.225 | 2.065 | 1.827 | 2.222 | 2.1 | 0.617 | 1.761 | 1.441 |
| 218.35 | 1.243 | 2.111 | 1.92 | 2.296 | 2.06 | 0.673 | 1.762 | 1.473 |
| 220.375 | 1.263 | 2.159 | 2.024 | 2.365 | 2.026 | 0.734 | 1.765 | 1.504 |
| 222.4 | 1.285 | 2.209 | 2.139 | 2.428 | 1.998 | 0.799 | 1.77 | 1.534 |
| 224.425 | 1.31 | 2.263 | 2.267 | 2.481 | 1.977 | 0.867 | 1.777 | 1.563 |
| 226.45 | 1.339 | 2.318 | 2.406 | 2.523 | 1.964 | 0.938 | 1.786 | 1.591 |
| 228.475 | 1.372 | 2.376 | 2.555 | 2.55 | 1.958 | 1.011 | 1.796 | 1.618 |
| 230.5 | 1.409 | 2.436 | 2.712 | 2.559 | 1.961 | 1.086 | 1.808 | 1.646 |
| 232.525 | 1.452 | 2.499 | 2.875 | 2.548 | 1.972 | 1.161 | 1.822 | 1.673 |
| 234.55 | 1.501 | 2.563 | 3.038 | 2.513 | 1.992 | 1.236 | 1.837 | 1.7 |
| 236.575 | 1.558 | 2.628 | 3.197 | 2.455 | 2.022 | 1.31 | 1.855 | 1.727 |
| 238.6 | 1.623 | 2.694 | 3.347 | 2.374 | 2.06 | 1.38 | 1.874 | 1.754 |
| 240.625 | 1.698 | 2.76 | 3.481 | 2.273 | 2.106 | 1.447 | 1.895 | 1.782 |
| 242.65 | 1.783 | 2.823 | 3.596 | 2.156 | 2.161 | 1.509 | 1.919 | 1.809 |
| 244.675 | 1.879 | 2.881 | 3.689 | 2.028 | 2.223 | 1.564 | 1.945 | 1.838 |
| 246.7 | 1.988 | 2.933 | 3.761 | 1.894 | 2.292 | 1.612 | 1.975 | 1.866 |
| 248.725 | 2.109 | 2.975 | 3.81 | 1.76 | 2.367 | 1.652 | 2.009 | 1.895 |
| 250.75 | 2.241 | 3.004 | 3.84 | 1.629 | 2.445 | 1.682 | 2.048 | 1.923 |
| 252.775 | 2.381 | 3.014 | 3.854 | 1.504 | 2.527 | 1.702 | 2.091 | 1.95 |
| 254.8 | 2.527 | 3.004 | 3.853 | 1.388 | 2.61 | 1.711 | 2.141 | 1.975 |
| 256.825 | 2.672 | 2.971 | 3.841 | 1.281 | 2.691 | 1.711 | 2.197 | 1.995 |
| 258.85 | 2.81 | 2.914 | 3.821 | 1.184 | 2.771 | 1.7 | 2.26 | 2.009 |
| 260.875 | 2.934 | 2.836 | 3.794 | 1.096 | 2.846 | 1.68 | 2.328 | 2.016 |
| 262.9 | 3.04 | 2.741 | 3.763 | 1.018 | 2.915 | 1.652 | 2.399 | 2.011 |
| 264.925 | 3.122 | 2.634 | 3.728 | 0.948 | 2.977 | 1.617 | 2.47 | 1.994 |
| 266.95 | 3.181 | 2.524 | 3.692 | 0.886 | 3.031 | 1.577 | 2.538 | 1.965 |
| 268.975 | 3.217 | 2.416 | 3.654 | 0.831 | 3.078 | 1.534 | 2.597 | 1.925 |
| 271 | 3.234 | 2.316 | 3.616 | 0.782 | 3.116 | 1.49 | 2.647 | 1.877 |
| 273.025 | 3.235 | 2.227 | 3.577 | 0.739 | 3.147 | 1.445 | 2.684 | 1.825 |
| 275.05 | 3.225 | 2.151 | 3.539 | 0.701 | 3.171 | 1.401 | 2.709 | 1.773 |
| 277.075 | 3.207 | 2.089 | 3.501 | 0.668 | 3.189 | 1.36 | 2.725 | 1.725 |
| 279.1 | 3.185 | 2.04 | 3.464 | 0.639 | 3.201 | 1.321 | 2.734 | 1.681 |
| 281.125 | 3.163 | 2.004 | 3.428 | 0.614 | 3.209 | 1.285 | 2.737 | 1.644 |
| 283.15 | 3.143 | 1.98 | 3.393 | 0.592 | 3.214 | 1.253 | 2.737 | 1.612 |
| 285.175 | 3.126 | 1.965 | 3.358 | 0.574 | 3.215 | 1.225 | 2.736 | 1.585 |
| 287.2 | 3.115 | 1.958 | 3.324 | 0.558 | 3.216 | 1.201 | 2.735 | 1.563 |
| 289.225 | 3.11 | 1.957 | 3.291 | 0.545 | 3.214 | 1.181 | 2.734 | 1.545 |
| 291.25 | 3.112 | 1.96 | 3.259 | 0.535 | 3.213 | 1.164 | 2.733 | 1.529 |
| 293.275 | 3.121 | 1.965 | 3.228 | 0.527 | 3.211 | 1.151 | 2.734 | 1.517 |
| 295.3 | 3.137 | 1.971 | 3.198 | 0.522 | 3.21 | 1.141 | 2.736 | 1.506 |
| 297.325 | 3.16 | 1.975 | 3.168 | 0.519 | 3.21 | 1.134 | 2.74 | 1.496 |
| 299.35 | 3.188 | 1.976 | 3.14 | 0.518 | 3.212 | 1.129 | 2.744 | 1.488 |
| 301.375 | 3.221 | 1.974 | 3.112 | 0.52 | 3.215 | 1.127 | 2.75 | 1.48 |
| 303.4 | 3.257 | 1.966 | 3.086 | 0.523 | 3.221 | 1.126 | 2.757 | 1.473 |
| 305.425 | 3.294 | 1.953 | 3.06 | 0.529 | 3.228 | 1.127 | 2.765 | 1.467 |
| 307.45 | 3.331 | 1.934 | 3.036 | 0.537 | 3.238 | 1.128 | 2.773 | 1.46 |
| 309.475 | 3.366 | 1.91 | 3.012 | 0.547 | 3.251 | 1.13 | 2.783 | 1.454 |
| 311.5 | 3.398 | 1.882 | 2.991 | 0.56 | 3.266 | 1.132 | 2.793 | 1.448 |
| 313.525 | 3.426 | 1.851 | 2.97 | 0.574 | 3.284 | 1.133 | 2.804 | 1.441 |
| 315.55 | 3.45 | 1.818 | 2.952 | 0.591 | 3.303 | 1.133 | 2.816 | 1.435 |
| 317.575 | 3.469 | 1.784 | 2.935 | 0.61 | 3.325 | 1.132 | 2.828 | 1.428 |
| 319.6 | 3.483 | 1.751 | 2.921 | 0.631 | 3.349 | 1.129 | 2.84 | 1.421 |
| 321.625 | 3.493 | 1.719 | 2.909 | 0.655 | 3.375 | 1.124 | 2.853 | 1.414 |
| 323.65 | 3.499 | 1.691 | 2.9 | 0.68 | 3.401 | 1.117 | 2.866 | 1.406 |
| 325.675 | 3.503 | 1.666 | 2.894 | 0.707 | 3.429 | 1.107 | 2.88 | 1.398 |
| 327.7 | 3.505 | 1.646 | 2.892 | 0.735 | 3.456 | 1.095 | 2.894 | 1.39 |
| 329.725 | 3.507 | 1.63 | 2.894 | 0.765 | 3.484 | 1.08 | 2.908 | 1.381 |
| 331.75 | 3.51 | 1.618 | 2.9 | 0.795 | 3.511 | 1.062 | 2.922 | 1.372 |
| 333.775 | 3.515 | 1.611 | 2.91 | 0.825 | 3.537 | 1.042 | 2.937 | 1.362 |
| 335.8 | 3.523 | 1.607 | 2.925 | 0.854 | 3.562 | 1.019 | 2.952 | 1.352 |
| 337.825 | 3.535 | 1.606 | 2.945 | 0.883 | 3.586 | 0.994 | 2.966 | 1.342 |
| 339.85 | 3.552 | 1.606 | 2.969 | 0.909 | 3.608 | 0.968 | 2.981 | 1.331 |
| 341.875 | 3.574 | 1.606 | 2.997 | 0.932 | 3.627 | 0.94 | 2.996 | 1.32 |
| 343.9 | 3.601 | 1.606 | 3.029 | 0.952 | 3.645 | 0.91 | 3.011 | 1.308 |
| 345.925 | 3.633 | 1.603 | 3.065 | 0.968 | 3.66 | 0.88 | 3.026 | 1.295 |
| 347.95 | 3.67 | 1.596 | 3.103 | 0.979 | 3.673 | 0.849 | 3.041 | 1.282 |
| 349.975 | 3.71 | 1.584 | 3.143 | 0.986 | 3.684 | 0.818 | 3.055 | 1.269 |
| 352 | 3.752 | 1.567 | 3.184 | 0.987 | 3.692 | 0.787 | 3.07 | 1.255 |
| 354.025 | 3.796 | 1.543 | 3.225 | 0.983 | 3.699 | 0.755 | 3.084 | 1.241 |
| 356.05 | 3.839 | 1.513 | 3.265 | 0.974 | 3.703 | 0.725 | 3.098 | 1.226 |
| 358.075 | 3.88 | 1.478 | 3.304 | 0.959 | 3.706 | 0.695 | 3.112 | 1.21 |
| 360.1 | 3.918 | 1.437 | 3.341 | 0.941 | 3.707 | 0.665 | 3.126 | 1.194 |
| 362.125 | 3.952 | 1.391 | 3.375 | 0.918 | 3.706 | 0.636 | 3.139 | 1.178 |
| 364.15 | 3.981 | 1.341 | 3.406 | 0.892 | 3.704 | 0.609 | 3.152 | 1.161 |
| 366.175 | 4.006 | 1.289 | 3.433 | 0.864 | 3.701 | 0.582 | 3.165 | 1.143 |
| 368.2 | 4.025 | 1.236 | 3.456 | 0.834 | 3.697 | 0.556 | 3.177 | 1.126 |
| 370.225 | 4.039 | 1.182 | 3.476 | 0.802 | 3.692 | 0.531 | 3.189 | 1.107 |
| 372.25 | 4.048 | 1.127 | 3.492 | 0.77 | 3.686 | 0.507 | 3.2 | 1.089 |
| 374.275 | 4.053 | 1.074 | 3.504 | 0.737 | 3.679 | 0.485 | 3.21 | 1.07 |
| 376.3 | 4.054 | 1.022 | 3.513 | 0.705 | 3.671 | 0.463 | 3.221 | 1.05 |
| 378.325 | 4.051 | 0.972 | 3.52 | 0.674 | 3.664 | 0.442 | 3.23 | 1.03 |
| 380.35 | 4.045 | 0.924 | 3.523 | 0.643 | 3.655 | 0.423 | 3.239 | 1.01 |
| 382.375 | 4.036 | 0.878 | 3.525 | 0.614 | 3.646 | 0.404 | 3.247 | 0.99 |
| 384.4 | 4.025 | 0.835 | 3.524 | 0.585 | 3.638 | 0.386 | 3.255 | 0.97 |
| 386.425 | 4.012 | 0.794 | 3.522 | 0.559 | 3.628 | 0.369 | 3.262 | 0.949 |
| 388.45 | 3.998 | 0.755 | 3.517 | 0.533 | 3.619 | 0.353 | 3.269 | 0.928 |
| 390.475 | 3.982 | 0.719 | 3.512 | 0.509 | 3.61 | 0.338 | 3.275 | 0.907 |
| 392.5 | 3.966 | 0.684 | 3.506 | 0.487 | 3.6 | 0.323 | 3.28 | 0.886 |
| 394.525 | 3.949 | 0.652 | 3.499 | 0.465 | 3.59 | 0.309 | 3.284 | 0.865 |
| 396.55 | 3.931 | 0.622 | 3.491 | 0.445 | 3.581 | 0.296 | 3.289 | 0.844 |
| 398.575 | 3.913 | 0.594 | 3.483 | 0.427 | 3.571 | 0.284 | 3.292 | 0.824 |
| 400.6 | 3.895 | 0.567 | 3.474 | 0.409 | 3.562 | 0.272 | 3.295 | 0.803 |
| 402.625 | 3.877 | 0.542 | 3.465 | 0.393 | 3.552 | 0.261 | 3.297 | 0.782 |
| 404.65 | 3.859 | 0.519 | 3.456 | 0.378 | 3.543 | 0.25 | 3.298 | 0.761 |
| 406.675 | 3.841 | 0.497 | 3.446 | 0.364 | 3.534 | 0.24 | 3.299 | 0.74 |
| 408.7 | 3.823 | 0.477 | 3.437 | 0.351 | 3.524 | 0.23 | 3.3 | 0.72 |
| 410.725 | 3.805 | 0.458 | 3.427 | 0.338 | 3.515 | 0.221 | 3.3 | 0.7 |
| 412.75 | 3.787 | 0.44 | 3.418 | 0.327 | 3.506 | 0.212 | 3.299 | 0.68 |
| 414.775 | 3.77 | 0.423 | 3.409 | 0.316 | 3.498 | 0.203 | 3.298 | 0.66 |
| 416.8 | 3.753 | 0.407 | 3.399 | 0.306 | 3.489 | 0.195 | 3.296 | 0.64 |
| 418.825 | 3.736 | 0.392 | 3.39 | 0.297 | 3.48 | 0.188 | 3.294 | 0.621 |
| 420.85 | 3.72 | 0.378 | 3.381 | 0.288 | 3.472 | 0.18 | 3.292 | 0.602 |
| 422.875 | 3.704 | 0.365 | 3.372 | 0.28 | 3.464 | 0.173 | 3.288 | 0.583 |
| 424.9 | 3.689 | 0.352 | 3.364 | 0.272 | 3.456 | 0.167 | 3.285 | 0.564 |
| 426.925 | 3.674 | 0.341 | 3.355 | 0.265 | 3.448 | 0.16 | 3.281 | 0.546 |
| 428.95 | 3.659 | 0.329 | 3.347 | 0.259 | 3.44 | 0.154 | 3.277 | 0.528 |
| 430.975 | 3.644 | 0.319 | 3.339 | 0.252 | 3.432 | 0.149 | 3.272 | 0.51 |
| 433 | 3.63 | 0.309 | 3.331 | 0.247 | 3.424 | 0.143 | 3.267 | 0.493 |
| 435.025 | 3.616 | 0.299 | 3.323 | 0.241 | 3.417 | 0.138 | 3.262 | 0.476 |
| 437.05 | 3.603 | 0.29 | 3.315 | 0.236 | 3.41 | 0.133 | 3.255 | 0.459 |
| 439.075 | 3.59 | 0.281 | 3.308 | 0.231 | 3.403 | 0.128 | 3.249 | 0.443 |
| 441.1 | 3.577 | 0.273 | 3.301 | 0.227 | 3.396 | 0.123 | 3.243 | 0.427 |
| 443.125 | 3.565 | 0.265 | 3.294 | 0.223 | 3.389 | 0.118 | 3.236 | 0.411 |
| 445.15 | 3.552 | 0.258 | 3.287 | 0.219 | 3.382 | 0.114 | 3.229 | 0.396 |
| 447.175 | 3.541 | 0.25 | 3.28 | 0.215 | 3.375 | 0.11 | 3.222 | 0.381 |
| 449.2 | 3.529 | 0.243 | 3.274 | 0.211 | 3.369 | 0.106 | 3.214 | 0.366 |
| 451.225 | 3.517 | 0.237 | 3.268 | 0.208 | 3.362 | 0.102 | 3.206 | 0.352 |
| 453.25 | 3.506 | 0.231 | 3.262 | 0.205 | 3.356 | 0.099 | 3.198 | 0.338 |
| 455.275 | 3.496 | 0.225 | 3.256 | 0.202 | 3.35 | 0.095 | 3.19 | 0.324 |
| 457.3 | 3.485 | 0.219 | 3.25 | 0.199 | 3.344 | 0.092 | 3.181 | 0.311 |
| 459.325 | 3.475 | 0.213 | 3.244 | 0.196 | 3.338 | 0.088 | 3.172 | 0.298 |
| 461.35 | 3.464 | 0.208 | 3.239 | 0.194 | 3.332 | 0.085 | 3.163 | 0.285 |
| 463.375 | 3.455 | 0.203 | 3.233 | 0.191 | 3.327 | 0.082 | 3.154 | 0.273 |
| 465.4 | 3.445 | 0.198 | 3.228 | 0.189 | 3.321 | 0.079 | 3.144 | 0.262 |
| 467.425 | 3.435 | 0.193 | 3.223 | 0.187 | 3.316 | 0.077 | 3.135 | 0.25 |
| 469.45 | 3.426 | 0.189 | 3.218 | 0.185 | 3.31 | 0.074 | 3.125 | 0.239 |
| 471.475 | 3.417 | 0.185 | 3.213 | 0.183 | 3.305 | 0.071 | 3.115 | 0.229 |
| 473.5 | 3.408 | 0.18 | 3.209 | 0.181 | 3.3 | 0.069 | 3.105 | 0.219 |
| 475.525 | 3.4 | 0.176 | 3.204 | 0.179 | 3.295 | 0.067 | 3.095 | 0.209 |
| 477.55 | 3.391 | 0.173 | 3.2 | 0.177 | 3.29 | 0.064 | 3.084 | 0.2 |
| 479.575 | 3.383 | 0.169 | 3.195 | 0.176 | 3.285 | 0.062 | 3.074 | 0.191 |
| 481.6 | 3.375 | 0.165 | 3.191 | 0.174 | 3.28 | 0.06 | 3.064 | 0.182 |
| 483.625 | 3.367 | 0.162 | 3.187 | 0.173 | 3.275 | 0.058 | 3.053 | 0.174 |
| 485.65 | 3.359 | 0.159 | 3.183 | 0.171 | 3.271 | 0.056 | 3.042 | 0.167 |
| 487.675 | 3.351 | 0.155 | 3.179 | 0.17 | 3.266 | 0.054 | 3.032 | 0.159 |
| 489.7 | 3.344 | 0.152 | 3.175 | 0.169 | 3.262 | 0.052 | 3.021 | 0.153 |
| 491.725 | 3.337 | 0.149 | 3.171 | 0.167 | 3.257 | 0.05 | 3.01 | 0.146 |
| 493.75 | 3.33 | 0.146 | 3.168 | 0.166 | 3.253 | 0.048 | 2.999 | 0.14 |
| 495.775 | 3.322 | 0.144 | 3.164 | 0.165 | 3.249 | 0.047 | 2.988 | 0.135 |
| 497.8 | 3.316 | 0.141 | 3.161 | 0.164 | 3.244 | 0.045 | 2.977 | 0.13 |
| 499.825 | 3.309 | 0.138 | 3.157 | 0.163 | 3.24 | 0.044 | 2.966 | 0.125 |
| 501.85 | 3.302 | 0.136 | 3.154 | 0.162 | 3.236 | 0.042 | 2.956 | 0.121 |
| 503.875 | 3.296 | 0.133 | 3.151 | 0.161 | 3.232 | 0.041 | 2.945 | 0.117 |
| 505.9 | 3.289 | 0.131 | 3.148 | 0.16 | 3.228 | 0.039 | 2.934 | 0.114 |
| 507.925 | 3.283 | 0.129 | 3.145 | 0.159 | 3.225 | 0.038 | 2.924 | 0.111 |
| 509.95 | 3.277 | 0.126 | 3.141 | 0.158 | 3.221 | 0.037 | 2.913 | 0.109 |
| 511.975 | 3.271 | 0.124 | 3.139 | 0.158 | 3.217 | 0.035 | 2.903 | 0.107 |
| 514 | 3.265 | 0.122 | 3.136 | 0.157 | 3.213 | 0.034 | 2.893 | 0.106 |
| 516.025 | 3.259 | 0.12 | 3.133 | 0.156 | 3.21 | 0.033 | 2.883 | 0.105 |
| 518.05 | 3.254 | 0.118 | 3.13 | 0.155 | 3.206 | 0.032 | 2.873 | 0.104 |
| 520.075 | 3.248 | 0.116 | 3.127 | 0.155 | 3.203 | 0.031 | 2.864 | 0.104 |
| 522.1 | 3.243 | 0.114 | 3.125 | 0.154 | 3.199 | 0.03 | 2.855 | 0.105 |
| 524.125 | 3.237 | 0.113 | 3.122 | 0.153 | 3.196 | 0.029 | 2.846 | 0.106 |
| 526.15 | 3.232 | 0.111 | 3.119 | 0.153 | 3.193 | 0.028 | 2.838 | 0.107 |
| 528.175 | 3.227 | 0.109 | 3.117 | 0.152 | 3.19 | 0.027 | 2.83 | 0.109 |
| 530.2 | 3.222 | 0.108 | 3.114 | 0.151 | 3.186 | 0.026 | 2.823 | 0.111 |
| 532.1 | 3.217 | 0.106 | 3.112 | 0.151 | 3.183 | 0.025 | 2.817 | 0.113 |
| 532.225 | 3.217 | 0.106 | 3.112 | 0.151 | 3.183 | 0.025 | 2.816 | 0.113 |
| 534.25 | 3.212 | 0.104 | 3.11 | 0.15 | 3.18 | 0.024 | 2.81 | 0.116 |
| 536.275 | 3.207 | 0.103 | 3.107 | 0.15 | 3.177 | 0.023 | 2.804 | 0.118 |
| 538.3 | 3.202 | 0.102 | 3.105 | 0.149 | 3.174 | 0.022 | 2.799 | 0.121 |
| 540.325 | 3.197 | 0.1 | 3.103 | 0.148 | 3.171 | 0.021 | 2.794 | 0.123 |
| 542.35 | 3.193 | 0.099 | 3.1 | 0.148 | 3.168 | 0.021 | 2.79 | 0.125 |
| 544.375 | 3.188 | 0.097 | 3.098 | 0.147 | 3.165 | 0.02 | 2.786 | 0.128 |
| 546.4 | 3.184 | 0.096 | 3.096 | 0.147 | 3.163 | 0.019 | 2.782 | 0.13 |
| 548.425 | 3.179 | 0.095 | 3.094 | 0.146 | 3.16 | 0.019 | 2.779 | 0.133 |
| 550.45 | 3.175 | 0.094 | 3.092 | 0.146 | 3.157 | 0.018 | 2.776 | 0.135 |
| 552.475 | 3.171 | 0.092 | 3.09 | 0.146 | 3.154 | 0.017 | 2.774 | 0.137 |
| 554.5 | 3.167 | 0.091 | 3.088 | 0.145 | 3.152 | 0.017 | 2.771 | 0.139 |
| 556.525 | 3.163 | 0.09 | 3.086 | 0.145 | 3.149 | 0.016 | 2.769 | 0.141 |
| 558.55 | 3.159 | 0.089 | 3.084 | 0.144 | 3.147 | 0.015 | 2.768 | 0.143 |
| 560.575 | 3.155 | 0.088 | 3.082 | 0.144 | 3.144 | 0.015 | 2.767 | 0.144 |
| 562.6 | 3.151 | 0.087 | 3.08 | 0.143 | 3.142 | 0.014 | 2.766 | 0.146 |
| 564.625 | 3.147 | 0.086 | 3.078 | 0.143 | 3.139 | 0.014 | 2.765 | 0.147 |
| 566.65 | 3.143 | 0.085 | 3.076 | 0.143 | 3.137 | 0.013 | 2.764 | 0.148 |
| 568.675 | 3.14 | 0.084 | 3.074 | 0.142 | 3.134 | 0.013 | 2.764 | 0.148 |
| 570.7 | 3.136 | 0.083 | 3.072 | 0.142 | 3.132 | 0.012 | 2.764 | 0.149 |
| 572.725 | 3.132 | 0.082 | 3.071 | 0.141 | 3.13 | 0.012 | 2.764 | 0.149 |
| 574.75 | 3.129 | 0.082 | 3.069 | 0.141 | 3.127 | 0.012 | 2.764 | 0.149 |
| 576.775 | 3.125 | 0.081 | 3.067 | 0.141 | 3.125 | 0.011 | 2.764 | 0.149 |
| 578.8 | 3.122 | 0.08 | 3.065 | 0.14 | 3.123 | 0.011 | 2.764 | 0.148 |
| 580.825 | 3.119 | 0.079 | 3.063 | 0.14 | 3.121 | 0.01 | 2.764 | 0.147 |
| 582.85 | 3.115 | 0.078 | 3.062 | 0.14 | 3.119 | 0.01 | 2.765 | 0.146 |
| 584.875 | 3.112 | 0.078 | 3.06 | 0.139 | 3.116 | 0.01 | 2.765 | 0.145 |
| 586.9 | 3.109 | 0.077 | 3.058 | 0.139 | 3.114 | 0.009 | 2.765 | 0.143 |
| 588.925 | 3.106 | 0.076 | 3.056 | 0.139 | 3.112 | 0.009 | 2.765 | 0.142 |
| 590.95 | 3.103 | 0.075 | 3.055 | 0.139 | 3.11 | 0.009 | 2.765 | 0.14 |
| 592.975 | 3.1 | 0.075 | 3.053 | 0.138 | 3.108 | 0.009 | 2.765 | 0.138 |
| 595 | 3.097 | 0.074 | 3.051 | 0.138 | 3.106 | 0.008 | 2.765 | 0.135 |
| 597.025 | 3.094 | 0.073 | 3.05 | 0.138 | 3.104 | 0.008 | 2.765 | 0.133 |
| 599.05 | 3.091 | 0.073 | 3.048 | 0.137 | 3.102 | 0.008 | 2.764 | 0.131 |
| 601.075 | 3.088 | 0.072 | 3.046 | 0.137 | 3.101 | 0.008 | 2.763 | 0.128 |
| 603.1 | 3.085 | 0.071 | 3.045 | 0.137 | 3.099 | 0.007 | 2.763 | 0.125 |
| 605.125 | 3.082 | 0.071 | 3.043 | 0.137 | 3.097 | 0.007 | 2.762 | 0.123 |
| 607.15 | 3.08 | 0.07 | 3.041 | 0.137 | 3.095 | 0.007 | 2.761 | 0.12 |
| 609.175 | 3.077 | 0.07 | 3.039 | 0.136 | 3.093 | 0.007 | 2.76 | 0.117 |
| 611.2 | 3.074 | 0.069 | 3.038 | 0.136 | 3.092 | 0.007 | 2.759 | 0.114 |
| 613.225 | 3.072 | 0.068 | 3.036 | 0.136 | 3.09 | 0.006 | 2.758 | 0.112 |
| 615.25 | 3.069 | 0.068 | 3.034 | 0.136 | 3.088 | 0.006 | 2.756 | 0.109 |
| 617.275 | 3.066 | 0.067 | 3.033 | 0.136 | 3.087 | 0.006 | 2.755 | 0.106 |
| 619.3 | 3.064 | 0.067 | 3.031 | 0.135 | 3.085 | 0.006 | 2.753 | 0.103 |
| 621.325 | 3.062 | 0.066 | 3.029 | 0.135 | 3.083 | 0.006 | 2.752 | 0.1 |
| 623.35 | 3.059 | 0.066 | 3.027 | 0.135 | 3.082 | 0.006 | 2.75 | 0.098 |
| 625.375 | 3.057 | 0.065 | 3.026 | 0.135 | 3.08 | 0.006 | 2.748 | 0.095 |
| 627.4 | 3.054 | 0.065 | 3.024 | 0.135 | 3.079 | 0.005 | 2.746 | 0.093 |
| 629.425 | 3.052 | 0.064 | 3.022 | 0.135 | 3.077 | 0.005 | 2.744 | 0.09 |
| 631.45 | 3.05 | 0.064 | 3.02 | 0.135 | 3.076 | 0.005 | 2.742 | 0.088 |
| 633.475 | 3.047 | 0.063 | 3.018 | 0.135 | 3.075 | 0.005 | 2.74 | 0.085 |
| 635.5 | 3.045 | 0.063 | 3.017 | 0.135 | 3.073 | 0.005 | 2.738 | 0.083 |
| 637.525 | 3.043 | 0.062 | 3.015 | 0.135 | 3.072 | 0.005 | 2.735 | 0.08 |
| 639.55 | 3.041 | 0.062 | 3.013 | 0.135 | 3.07 | 0.005 | 2.733 | 0.078 |
| 641.575 | 3.039 | 0.061 | 3.011 | 0.135 | 3.069 | 0.005 | 2.731 | 0.076 |
| 643.6 | 3.037 | 0.061 | 3.009 | 0.135 | 3.068 | 0.005 | 2.728 | 0.074 |
| 645.625 | 3.034 | 0.061 | 3.007 | 0.136 | 3.066 | 0.004 | 2.726 | 0.072 |
| 647.65 | 3.032 | 0.06 | 3.005 | 0.136 | 3.065 | 0.004 | 2.724 | 0.07 |
| 649.675 | 3.03 | 0.06 | 3.003 | 0.136 | 3.064 | 0.004 | 2.721 | 0.068 |
| 651.7 | 3.028 | 0.059 | 3.001 | 0.136 | 3.062 | 0.004 | 2.719 | 0.066 |
| 653.725 | 3.026 | 0.059 | 2.998 | 0.137 | 3.061 | 0.004 | 2.717 | 0.064 |
| 655.75 | 3.025 | 0.059 | 2.996 | 0.137 | 3.06 | 0.004 | 2.714 | 0.063 |
| 657.775 | 3.023 | 0.058 | 2.994 | 0.137 | 3.059 | 0.004 | 2.712 | 0.061 |
| 659.8 | 3.021 | 0.058 | 2.992 | 0.138 | 3.058 | 0.004 | 2.709 | 0.059 |
| 661.825 | 3.019 | 0.057 | 2.989 | 0.138 | 3.056 | 0.004 | 2.707 | 0.058 |
| 663.85 | 3.017 | 0.057 | 2.987 | 0.139 | 3.055 | 0.003 | 2.705 | 0.056 |
| 665.875 | 3.015 | 0.057 | 2.984 | 0.14 | 3.054 | 0.003 | 2.702 | 0.055 |
| 667.9 | 3.013 | 0.056 | 2.982 | 0.141 | 3.053 | 0.003 | 2.7 | 0.053 |
| 669.925 | 3.011 | 0.056 | 2.979 | 0.142 | 3.052 | 0.003 | 2.698 | 0.052 |
| 671.95 | 3.01 | 0.055 | 2.977 | 0.143 | 3.051 | 0.003 | 2.695 | 0.051 |
| 673.975 | 3.008 | 0.055 | 2.974 | 0.144 | 3.05 | 0.003 | 2.693 | 0.049 |
| 676 | 3.006 | 0.055 | 2.971 | 0.145 | 3.048 | 0.003 | 2.691 | 0.048 |
| 678.025 | 3.004 | 0.054 | 2.968 | 0.146 | 3.047 | 0.003 | 2.688 | 0.047 |
| 680.05 | 3.003 | 0.054 | 2.965 | 0.148 | 3.046 | 0.003 | 2.686 | 0.046 |
| 682.075 | 3.001 | 0.054 | 2.962 | 0.15 | 3.045 | 0.003 | 2.684 | 0.044 |
| 684.1 | 2.999 | 0.053 | 2.959 | 0.152 | 3.044 | 0.003 | 2.682 | 0.043 |
| 686.125 | 2.998 | 0.053 | 2.955 | 0.154 | 3.043 | 0.003 | 2.68 | 0.042 |
| 688.15 | 2.996 | 0.053 | 2.952 | 0.156 | 3.042 | 0.002 | 2.677 | 0.041 |
| 690.175 | 2.995 | 0.052 | 2.948 | 0.159 | 3.041 | 0.002 | 2.675 | 0.04 |
| 692.2 | 2.993 | 0.052 | 2.945 | 0.162 | 3.04 | 0.002 | 2.673 | 0.039 |
| 694.225 | 2.992 | 0.052 | 2.941 | 0.166 | 3.039 | 0.002 | 2.671 | 0.038 |
| 696.25 | 2.99 | 0.051 | 2.937 | 0.17 | 3.038 | 0.002 | 2.669 | 0.037 |
| 698.275 | 2.989 | 0.051 | 2.934 | 0.174 | 3.037 | 0.002 | 2.667 | 0.037 |
| 700.3 | 2.987 | 0.051 | 2.93 | 0.179 | 3.036 | 0.002 | 2.664 | 0.036 |
| 702.325 | 2.986 | 0.051 | 2.926 | 0.184 | 3.035 | 0.002 | 2.662 | 0.035 |
| 704.35 | 2.984 | 0.05 | 2.922 | 0.19 | 3.034 | 0.002 | 2.66 | 0.034 |
| 706.375 | 2.983 | 0.05 | 2.919 | 0.197 | 3.034 | 0.002 | 2.658 | 0.033 |
| 708.4 | 2.981 | 0.05 | 2.916 | 0.205 | 3.033 | 0.002 | 2.656 | 0.033 |
| 710.425 | 2.98 | 0.049 | 2.913 | 0.214 | 3.032 | 0.002 | 2.654 | 0.032 |
| 712.45 | 2.978 | 0.049 | 2.91 | 0.223 | 3.031 | 0.002 | 2.653 | 0.031 |
| 714.475 | 2.977 | 0.049 | 2.909 | 0.234 | 3.03 | 0.002 | 2.651 | 0.03 |
| 716.5 | 2.976 | 0.049 | 2.908 | 0.246 | 3.029 | 0.002 | 2.649 | 0.03 |
| 718.525 | 2.974 | 0.048 | 2.909 | 0.259 | 3.028 | 0.002 | 2.647 | 0.029 |
| 720.55 | 2.973 | 0.048 | 2.912 | 0.273 | 3.027 | 0.001 | 2.645 | 0.028 |
| 722.575 | 2.971 | 0.048 | 2.916 | 0.287 | 3.027 | 0.001 | 2.643 | 0.028 |
| 724.6 | 2.97 | 0.047 | 2.923 | 0.302 | 3.026 | 0.001 | 2.641 | 0.027 |
| 726.625 | 2.969 | 0.047 | 2.933 | 0.317 | 3.025 | 0.001 | 2.639 | 0.027 |
| 728.65 | 2.968 | 0.047 | 2.946 | 0.331 | 3.024 | 0.001 | 2.638 | 0.026 |
| 730.675 | 2.966 | 0.047 | 2.963 | 0.343 | 3.023 | 0.001 | 2.636 | 0.026 |
| 732.7 | 2.965 | 0.046 | 2.981 | 0.353 | 3.023 | 0.001 | 2.634 | 0.025 |
| 734.725 | 2.964 | 0.046 | 3.002 | 0.359 | 3.022 | 0.001 | 2.632 | 0.025 |
| 736.75 | 2.963 | 0.046 | 3.025 | 0.362 | 3.021 | 0.001 | 2.631 | 0.024 |
| 738.775 | 2.961 | 0.046 | 3.047 | 0.36 | 3.02 | 0.001 | 2.629 | 0.024 |
| 740.8 | 2.96 | 0.045 | 3.068 | 0.354 | 3.019 | 0.001 | 2.627 | 0.023 |
| 742.825 | 2.959 | 0.045 | 3.088 | 0.345 | 3.019 | 0.001 | 2.625 | 0.023 |
| 744.85 | 2.958 | 0.045 | 3.104 | 0.333 | 3.018 | 0.001 | 2.624 | 0.022 |
| 746.875 | 2.957 | 0.045 | 3.118 | 0.319 | 3.017 | 0.001 | 2.622 | 0.022 |
| 748.9 | 2.955 | 0.045 | 3.129 | 0.304 | 3.016 | 0.001 | 2.621 | 0.021 |
| 750.925 | 2.954 | 0.044 | 3.137 | 0.289 | 3.016 | 0.001 | 2.619 | 0.021 |
| 752.95 | 2.953 | 0.044 | 3.142 | 0.274 | 3.015 | 0.001 | 2.617 | 0.02 |
| 754.975 | 2.952 | 0.044 | 3.145 | 0.26 | 3.014 | 0.001 | 2.616 | 0.02 |
| 757 | 2.951 | 0.044 | 3.147 | 0.247 | 3.014 | 0.001 | 2.614 | 0.02 |
| 759.025 | 2.95 | 0.043 | 3.147 | 0.235 | 3.013 | 0.001 | 2.613 | 0.019 |
| 761.05 | 2.949 | 0.043 | 3.146 | 0.223 | 3.012 | 0.001 | 2.611 | 0.019 |
| 763.075 | 2.948 | 0.043 | 3.144 | 0.213 | 3.011 | 0.001 | 2.61 | 0.019 |
| 765.1 | 2.947 | 0.043 | 3.142 | 0.204 | 3.011 | 0.001 | 2.608 | 0.018 |
| 767.125 | 2.945 | 0.043 | 3.139 | 0.195 | 3.01 | 0.001 | 2.607 | 0.018 |
| 769.15 | 2.944 | 0.042 | 3.136 | 0.188 | 3.009 | 0.001 | 2.605 | 0.018 |
| 771.175 | 2.943 | 0.042 | 3.132 | 0.181 | 3.009 | 0 | 2.604 | 0.017 |
| 773.2 | 2.942 | 0.042 | 3.129 | 0.174 | 3.008 | 0 | 2.603 | 0.017 |
| 775.225 | 2.941 | 0.042 | 3.125 | 0.169 | 3.007 | 0 | 2.601 | 0.017 |
| 777.25 | 2.94 | 0.042 | 3.121 | 0.164 | 3.007 | 0 | 2.6 | 0.016 |
| 779.275 | 2.939 | 0.041 | 3.118 | 0.159 | 3.006 | 0 | 2.599 | 0.016 |
| 781.3 | 2.938 | 0.041 | 3.114 | 0.155 | 3.006 | 0 | 2.597 | 0.016 |
| 783.325 | 2.937 | 0.041 | 3.111 | 0.151 | 3.005 | 0 | 2.596 | 0.015 |
| 785.35 | 2.936 | 0.041 | 3.107 | 0.148 | 3.004 | 0 | 2.595 | 0.015 |
| 787.375 | 2.935 | 0.041 | 3.104 | 0.145 | 3.004 | 0 | 2.593 | 0.015 |
| 789.4 | 2.934 | 0.04 | 3.1 | 0.142 | 3.003 | 0 | 2.592 | 0.015 |
| 791.425 | 2.933 | 0.04 | 3.097 | 0.139 | 3.003 | 0 | 2.591 | 0.014 |
| 793.45 | 2.933 | 0.04 | 3.094 | 0.137 | 3.002 | 0 | 2.589 | 0.014 |
| 795.475 | 2.932 | 0.04 | 3.091 | 0.135 | 3.001 | 0 | 2.588 | 0.014 |
| 797.5 | 2.931 | 0.04 | 3.088 | 0.133 | 3.001 | 0 | 2.587 | 0.014 |
| 799.525 | 2.93 | 0.039 | 3.085 | 0.131 | 3 | 0 | 2.586 | 0.013 |
| 801.55 | 2.929 | 0.039 | 3.082 | 0.129 | 3 | 0 | 2.584 | 0.013 |
| 803.575 | 2.928 | 0.039 | 3.079 | 0.127 | 2.999 | 0 | 2.583 | 0.013 |
| 805.6 | 2.927 | 0.039 | 3.076 | 0.126 | 2.999 | 0 | 2.582 | 0.013 |
| 807.625 | 2.926 | 0.039 | 3.074 | 0.124 | 2.998 | 0 | 2.581 | 0.013 |
| 809.65 | 2.925 | 0.039 | 3.071 | 0.123 | 2.997 | 0 | 2.58 | 0.012 |
| 811.675 | 2.924 | 0.038 | 3.068 | 0.122 | 2.997 | 0 | 2.579 | 0.012 |
| 813.7 | 2.924 | 0.038 | 3.066 | 0.121 | 2.996 | 9.28E-05 | 2.577 | 0.012 |
| 815.725 | 2.923 | 0.038 | 3.063 | 0.12 | 2.996 | 8.21E-05 | 2.576 | 0.012 |
| 817.75 | 2.922 | 0.038 | 3.061 | 0.119 | 2.995 | 7.21E-05 | 2.575 | 0.012 |
| 819.775 | 2.921 | 0.038 | 3.058 | 0.118 | 2.995 | 6.27E-05 | 2.574 | 0.011 |
| 821.8 | 2.92 | 0.038 | 3.056 | 0.117 | 2.994 | 5.41E-05 | 2.573 | 0.011 |
| 823.825 | 2.919 | 0.037 | 3.053 | 0.116 | 2.994 | 4.61E-05 | 2.572 | 0.011 |
| 825.85 | 2.919 | 0.037 | 3.051 | 0.116 | 2.993 | 3.87E-05 | 2.571 | 0.011 |
| 827.875 | 2.918 | 0.037 | 3.049 | 0.115 | 2.993 | 3.21E-05 | 2.57 | 0.011 |
| 829.9 | 2.917 | 0.037 | 3.046 | 0.114 | 2.992 | 2.60E-05 | 2.569 | 0.011 |
| 831.925 | 2.916 | 0.037 | 3.044 | 0.114 | 2.992 | 2.07E-05 | 2.567 | 0.01 |
| 833.95 | 2.915 | 0.037 | 3.042 | 0.113 | 2.991 | 1.59E-05 | 2.566 | 0.01 |
| 835.975 | 2.915 | 0.036 | 3.039 | 0.113 | 2.991 | 1.18E-05 | 2.565 | 0.01 |
| 838 | 2.914 | 0.036 | 3.037 | 0.113 | 2.99 | 8.30E-06 | 2.564 | 0.01 |
| 840.025 | 2.913 | 0.036 | 3.035 | 0.112 | 2.99 | 5.43E-06 | 2.563 | 0.01 |
| 842.05 | 2.912 | 0.036 | 3.032 | 0.112 | 2.989 | 3.18E-06 | 2.562 | 0.01 |
| 844.075 | 2.912 | 0.036 | 3.03 | 0.112 | 2.989 | 1.53E-06 | 2.561 | 0.009 |
| 846.1 | 2.911 | 0.036 | 3.028 | 0.111 | 2.988 | 4.76E-07 | 2.56 | 0.009 |
| 848.125 | 2.91 | 0.036 | 3.025 | 0.111 | 2.988 | 2.15E-08 | 2.56 | 0.009 |
| 850.15 | 2.909 | 0.035 | 3.023 | 0.111 | 2.987 | 0 | 2.559 | 0.009 |
| 852.175 | 2.909 | 0.035 | 3.021 | 0.111 | 2.987 | 0 | 2.558 | 0.009 |
| 854.2 | 2.908 | 0.035 | 3.019 | 0.111 | 2.987 | 0 | 2.557 | 0.009 |
| 856.225 | 2.907 | 0.035 | 3.016 | 0.111 | 2.986 | 0 | 2.556 | 0.009 |
| 858.25 | 2.906 | 0.035 | 3.014 | 0.111 | 2.986 | 0 | 2.555 | 0.009 |
| 860.275 | 2.906 | 0.035 | 3.012 | 0.111 | 2.985 | 0 | 2.554 | 0.008 |
| 862.3 | 2.905 | 0.035 | 3.009 | 0.111 | 2.985 | 0 | 2.553 | 0.008 |
| 864.325 | 2.904 | 0.034 | 3.007 | 0.112 | 2.984 | 0 | 2.552 | 0.008 |
| 866.35 | 2.904 | 0.034 | 3.004 | 0.112 | 2.984 | 0 | 2.551 | 0.008 |
| 868.375 | 2.903 | 0.034 | 3.002 | 0.112 | 2.983 | 0 | 2.55 | 0.008 |
| 870.4 | 2.902 | 0.034 | 2.999 | 0.113 | 2.983 | 0 | 2.549 | 0.008 |
| 872.425 | 2.902 | 0.034 | 2.997 | 0.113 | 2.983 | 0 | 2.548 | 0.008 |
| 874.45 | 2.901 | 0.034 | 2.994 | 0.114 | 2.982 | 0 | 2.548 | 0.008 |
| 876.475 | 2.9 | 0.034 | 2.992 | 0.114 | 2.982 | 0 | 2.547 | 0.007 |
| 878.5 | 2.9 | 0.033 | 2.989 | 0.115 | 2.981 | 0 | 2.546 | 0.007 |
| 880.525 | 2.899 | 0.033 | 2.986 | 0.115 | 2.981 | 0 | 2.545 | 0.007 |
| 882.55 | 2.898 | 0.033 | 2.984 | 0.116 | 2.981 | 0 | 2.544 | 0.007 |
| 884.575 | 2.898 | 0.033 | 2.981 | 0.117 | 2.98 | 0 | 2.543 | 0.007 |
| 886.6 | 2.897 | 0.033 | 2.978 | 0.118 | 2.98 | 0 | 2.543 | 0.007 |
| 888.625 | 2.896 | 0.033 | 2.975 | 0.119 | 2.98 | 0 | 2.542 | 0.007 |
| 890.65 | 2.896 | 0.033 | 2.972 | 0.12 | 2.979 | 0 | 2.541 | 0.007 |
| 892.675 | 2.895 | 0.033 | 2.969 | 0.121 | 2.979 | 0 | 2.54 | 0.007 |
| 894.7 | 2.895 | 0.032 | 2.966 | 0.123 | 2.978 | 0 | 2.539 | 0.007 |
| 896.725 | 2.894 | 0.032 | 2.963 | 0.124 | 2.978 | 0 | 2.539 | 0.006 |
| 898.75 | 2.893 | 0.032 | 2.96 | 0.126 | 2.978 | 0 | 2.538 | 0.006 |
| 900.775 | 2.893 | 0.032 | 2.957 | 0.128 | 2.977 | 0 | 2.537 | 0.006 |
| 902.8 | 2.892 | 0.032 | 2.953 | 0.13 | 2.977 | 0 | 2.536 | 0.006 |
| 904.825 | 2.892 | 0.032 | 2.95 | 0.132 | 2.977 | 0 | 2.536 | 0.006 |
| 906.85 | 2.891 | 0.032 | 2.947 | 0.134 | 2.976 | 0 | 2.535 | 0.006 |
| 908.875 | 2.89 | 0.032 | 2.943 | 0.136 | 2.976 | 0 | 2.534 | 0.006 |
| 910.9 | 2.89 | 0.031 | 2.939 | 0.139 | 2.976 | 0 | 2.533 | 0.006 |
| 912.925 | 2.889 | 0.031 | 2.936 | 0.142 | 2.975 | 0 | 2.533 | 0.006 |
| 914.95 | 2.889 | 0.031 | 2.932 | 0.145 | 2.975 | 0 | 2.532 | 0.006 |
| 916.975 | 2.888 | 0.031 | 2.928 | 0.148 | 2.974 | 0 | 2.531 | 0.006 |
| 919 | 2.888 | 0.031 | 2.924 | 0.152 | 2.974 | 0 | 2.53 | 0.006 |
| 921.025 | 2.887 | 0.031 | 2.92 | 0.156 | 2.974 | 0 | 2.53 | 0.005 |
| 923.05 | 2.886 | 0.031 | 2.916 | 0.16 | 2.973 | 0 | 2.529 | 0.005 |
| 925.075 | 2.886 | 0.031 | 2.912 | 0.165 | 2.973 | 0 | 2.528 | 0.005 |
| 927.1 | 2.885 | 0.031 | 2.908 | 0.17 | 2.973 | 0 | 2.528 | 0.005 |
| 929.125 | 2.885 | 0.03 | 2.904 | 0.175 | 2.973 | 0 | 2.527 | 0.005 |
| 931.15 | 2.884 | 0.03 | 2.9 | 0.181 | 2.972 | 0 | 2.526 | 0.005 |
| 933.175 | 2.884 | 0.03 | 2.896 | 0.188 | 2.972 | 0 | 2.526 | 0.005 |
| 935.2 | 2.883 | 0.03 | 2.893 | 0.195 | 2.972 | 0 | 2.525 | 0.005 |
| 937.225 | 2.883 | 0.03 | 2.889 | 0.203 | 2.971 | 0 | 2.524 | 0.005 |
| 939.25 | 2.882 | 0.03 | 2.885 | 0.211 | 2.971 | 0 | 2.524 | 0.005 |
| 941.275 | 2.882 | 0.03 | 2.882 | 0.22 | 2.971 | 0 | 2.523 | 0.005 |
| 943.3 | 2.881 | 0.03 | 2.879 | 0.23 | 2.97 | 0 | 2.522 | 0.005 |
| 945.325 | 2.881 | 0.03 | 2.876 | 0.24 | 2.97 | 0 | 2.522 | 0.005 |
| 947.35 | 2.88 | 0.029 | 2.874 | 0.252 | 2.97 | 0 | 2.521 | 0.005 |
| 949.375 | 2.88 | 0.029 | 2.873 | 0.264 | 2.969 | 0 | 2.52 | 0.004 |
| 951.4 | 2.879 | 0.029 | 2.872 | 0.277 | 2.969 | 0 | 2.52 | 0.004 |
| 953.425 | 2.879 | 0.029 | 2.872 | 0.291 | 2.969 | 0 | 2.519 | 0.004 |
| 955.45 | 2.878 | 0.029 | 2.874 | 0.306 | 2.968 | 0 | 2.518 | 0.004 |
| 957.475 | 2.878 | 0.029 | 2.877 | 0.321 | 2.968 | 0 | 2.518 | 0.004 |
| 959.5 | 2.877 | 0.029 | 2.881 | 0.337 | 2.968 | 0 | 2.517 | 0.004 |
| 961.525 | 2.877 | 0.029 | 2.887 | 0.354 | 2.968 | 0 | 2.517 | 0.004 |
| 963.55 | 2.876 | 0.029 | 2.895 | 0.371 | 2.967 | 0 | 2.516 | 0.004 |
| 965.575 | 2.876 | 0.029 | 2.906 | 0.388 | 2.967 | 0 | 2.515 | 0.004 |
| 967.6 | 2.875 | 0.028 | 2.919 | 0.405 | 2.967 | 0 | 2.515 | 0.004 |
| 969.625 | 2.875 | 0.028 | 2.934 | 0.421 | 2.966 | 0 | 2.514 | 0.004 |
| 971.65 | 2.874 | 0.028 | 2.951 | 0.436 | 2.966 | 0 | 2.514 | 0.004 |
| 973.675 | 2.874 | 0.028 | 2.971 | 0.448 | 2.966 | 0 | 2.513 | 0.004 |
| 975.7 | 2.873 | 0.028 | 2.993 | 0.459 | 2.966 | 0 | 2.513 | 0.004 |
| 977.725 | 2.873 | 0.028 | 3.017 | 0.467 | 2.965 | 0 | 2.512 | 0.004 |
| 979.75 | 2.873 | 0.028 | 3.042 | 0.472 | 2.965 | 0 | 2.511 | 0.004 |
| 981.775 | 2.872 | 0.028 | 3.068 | 0.474 | 2.965 | 0 | 2.511 | 0.004 |
| 983.8 | 2.872 | 0.028 | 3.093 | 0.473 | 2.965 | 0 | 2.51 | 0.004 |
| 985.825 | 2.871 | 0.028 | 3.118 | 0.468 | 2.964 | 0 | 2.51 | 0.004 |
| 987.85 | 2.871 | 0.028 | 3.142 | 0.461 | 2.964 | 0 | 2.509 | 0.003 |
| 989.875 | 2.87 | 0.027 | 3.165 | 0.45 | 2.964 | 0 | 2.509 | 0.003 |
| 991.9 | 2.87 | 0.027 | 3.185 | 0.438 | 2.964 | 0 | 2.508 | 0.003 |
| 993.925 | 2.869 | 0.027 | 3.203 | 0.424 | 2.963 | 0 | 2.508 | 0.003 |
| 995.95 | 2.869 | 0.027 | 3.219 | 0.408 | 2.963 | 0 | 2.507 | 0.003 |
| 997.975 | 2.869 | 0.027 | 3.232 | 0.392 | 2.963 | 0 | 2.507 | 0.003 |
| 1000 | 2.868 | 0.027 | 3.243 | 0.375 | 2.963 | 0 | 2.506 | 0.003 |